%% file: main.tex
\documentclass[conference]{IEEEtran}
\IEEEoverridecommandlockouts

\usepackage{dblfloatfix}
\usepackage{lipsum}
\usepackage{amsmath,amssymb,amsfonts}
\usepackage{algorithmic}
\usepackage{graphicx}
\usepackage{dblfloatfix}
\usepackage{textcomp}   
\usepackage{xcolor}
\usepackage{float}
\usepackage{multirow}
\usepackage{footnote}
\usepackage{enumerate}
\usepackage{comment}

\usepackage{subcaption}

\usepackage{booktabs}

\usepackage{acronym}
\input{./chapter/acronyms.tex}

\input{chapter/commands.tex}
\usepackage[
    pdfborder={0 0 0},
    colorlinks=false,
    urlcolor=black,
    linkcolor=black,
    citecolor=black,
    filecolor=black,
    breaklinks,
]{hyperref}

\def\BibTeX{{\rm B\kern-.05em{\sc i\kern-.025em b}\kern-.08em
    T\kern-.1667em\lower.7ex\hbox{E}\kern-.125emX}}
    
\begin{document}

\title{Latency optimized Deep Neural Networks (DNNs): An Artificial Intelligence approach at the Edge using Multiprocessor System on Chip (MPSoC)}

\author{
\IEEEauthorblockN{Seyed Nima Omidsajedi\IEEEauthorrefmark{2}, Rekha Reddy\IEEEauthorrefmark{2}, Jianming Yi\IEEEauthorrefmark{2}, Jan Herbst\IEEEauthorrefmark{2}, Christoph Lipps\IEEEauthorrefmark{2} and Hans Dieter Schotten\IEEEauthorrefmark{1}\IEEEauthorrefmark{2}}
\IEEEauthorblockA{\IEEEauthorrefmark{2}Intelligent Networks Research
    Group, German Research Center for Artificial Intelligence\\ D-67663
    Kaiserslautern, Email: \{firstname.lastname\}@dfki.de}
\IEEEauthorblockA{\IEEEauthorrefmark{1}Institute for Wireless
    Communication and Navigation, University of Kaiserslautern\\ D-67663
    Kaiserslautern, mail: \{lastname\}@eit.uni-kl.de}
}

\maketitle
\pagestyle{plain}

\begin{abstract}
Almost in every heavily computation-dependent application, from 6G communication systems to autonomous driving platforms, a large portion of computing should be near to the client side. Edge computing (AI at Edge) in mobile devices is one of the optimized approaches for addressing this requirement. 

Therefore, in this work, the possibilities and challenges of implementing a low-latency and power-optimized smart mobile system are examined. Utilizing \ac{fpga} based solutions at the edge will lead to bandwidth-optimized designs and as a consequence can boost the computational effectiveness at a system-level deadline. Moreover, various performance aspects and implementation feasibilities of Neural Networks (NNs) on both embedded \ac{fpga} edge devices (using Xilinx\ac{mpsoc}) and Cloud are discussed throughout this research. The main goal of this work is to demonstrate a hybrid system that uses the deep learning programmable engine developed by Xilinx Inc. as the main component of the hardware accelerator. Then based on this design, an efficient system for mobile edge computing is represented by utilizing an embedded solution.
\end{abstract}


\section{Introduction: Edge computing platforms} 
\input{./chapter/introduction.tex}

\section{Related works: Using Embedded FPGAs as Edge computing systems}
\input{./chapter/related_works.tex}

\section{Proposed system: Hardware accelerator on embedded FPGA for DNNs}
\input{./chapter/our_work.tex}

\section{Performance results of DNN: On Edge versus on Cloud}
\input{./chapter/comparison_edge_cloud.tex}

\section{Conclusion \& Future Work}
\input{./chapter/conclusion.tex}

\bibliographystyle{IEEEtran}
\bibliography{main}

\end{document}

%% file: chapter/acronyms.tex
\newacro{5g}[5G]{Fifth Generation}
\newacro{6g}[6G]{Sixth Generation}
\newacro{as}[AS]{Authentication Server}
\newacro{atm}[ATM]{Automated Teller Machine}
\newacro{ai}[AI]{Artificial Intelligence}
\newacro{arm}[ARM]{Advanced RISC Machines}
\newacro{anna}[ANNA]{Analog Neural Network Arithmetic and logic unit}
\newacro{b5g}[B5G]{Beyond 5G}
\newacro{ban}[BAN]{Body Area Network}
\newacro{bsi}[\textit{BSI}]{\textit{Federal Office for Information Security}}
\newacro{bdr}[BDR]{Bit Disagreement Rate}
\newacro{bs}[BS]{Base Station}
\newacro{bw}[BW]{Band-Width}
\newacro{bram}[BRAM]{Block Random Access Memory}
\newacro{ca}[CA]{Certification Authority}
\newacro{cav}[CAV]{Connected Autonomous Vehicles}
\newacro{cc}[CC]{Common Criteria}
\newacro{ccu}[CCU]{Central Control Unit}
\newacro{cir}[CIR]{Channel Impulse Response}
\newacro{cr}[CR]{Challenge-Response}
\newacro{cpu}[CPU]{Central Processing Unit}
\newacro{cpps}[CPPS]{Cyber-Physical Production System}
\newacro{crl}[CRL]{Certificate Revocation List}
\newacro{csi}[CSI]{Channel State Information}
\newacro{crke}[CRKE]{Channel-Reciprocity Based Key Extraction}
\newacro{ctf}[CTF]{Channel Transfer Function}
\newacro{cotf}[COTF]{Commercial-off-the-Shelf}
\newacro{cmos}[CMOS]{Complementary Metal-Oxide-Semiconductors}
\newacro{dnn}[DNN]{Deep Neural Network}
\newacro{dos}[DoS]{Denial-of-Service}
\newacro{ddos}[DDoS]{Distributed-Denial-of-Service}
\newacro{dna}[DNA]{Deoxyribonucleic Acid}
\newacro{dtls}[DTLS]{Datagram Transport Layer Security}
\newacro{dct}[DCT]{Discrete Cosine Transformation}
\newacro{dlt}[DLT]{Distributed Ledger Technology}
\newacro{dpu}[DPU]{Deep Learning Processing Unit}
\newacro{dsp}[DSP]{Digital Signal Processor}

\newacro{eal}[EAL]{Evaluation Assurance Level}
\newacro{ecc}[ECC]{Elliptic Curve Cryptography}
\newacro{ecg}[ECG]{Electrocardiogram}
\newacro{eeg}[EEG]{Electroencephalogram}
\newacro{embb}[eMBB]{enhanced Mobile Broad-Band}
\newacro{emg}[EMG]{Electromyogram}
\newacro{eog}[EOG]{Electrooculography}
\newacro{enb}[eNodeB]{Evolved Node B}
\newacro{er}[ER]{Extended Reality}
\newacro{fpga}[FPGA]{Field Programmable Gate Array}
\newacro{fdd}[FDD]{Frequency Division Duplexing}
\newacro{ff}[FF]{Flip-Flop}
\newacro{gdpr}[GDPR]{General Data Protection Regulation}
\newacro{gd}[G\&D]{Giesecke \& Devrient}
\newacro{gpu}[GPU]{Graphics Processing Unit}
\newacro{h2m}[H2M]{Human-to-Machine}
\newacro{h2s}[H2S]{Human-to-Service}
\newacro{hmac}[HMAC]{Keyed-Hash Message Authentication Code}
\newacro{htc}[HTC]{Hologaphic-Type Communication}
\newacro{hotp}[HOTP]{HMAC-based One-time Password Algorithm}
\newacro{hsm}[HSM]{Hardware Security Module}
\newacro{iacs}[IACS]{Industrial Automation and Control System}
\newacro{ic}[IC]{Integrated Circuit}
\newacro{ics}[ICS]{Industrial Control System}
\newacro{id}[ID]{Identificator}
\newacro{ids}[IDS]{Intursion Detection System}
\newacro{iiot}[IIoT]{Industrial Internet of Things}
\newacro{io}[I/O]{Input/Output}
\newacro{ioe}[IoE]{Internet of Everything}
\newacro{iot}[IoT]{Internet of Things}
\newacro{irs}[IRS]{Intelligent Reflecting Surface}
\newacro{istn}[ISTN]{Integrated Space and Terrestrial Network}
\newacro{it}[IT]{Information Technology}
\newacro{itu}[ITU]{International Telecommunication Union}
\newacro{jcop}[JCOP]{Java Card Open Platform}
\newacro{kba}[KBA]{Knowledge Based Authentication}
\newacro{kdf}[KDF]{Key Derivation Function}
\newacro{lte}[LTE]{Long Term Evolution}
\newacro{ltea}[LTE-A]{Long Term Evolution Advanced}
\newacro{lr}[LR]{Linear Regression}
\newacro{los}[LoS]{Line of Sight}
\newacro{lorawan}[LoRaWAN]{Long Range Wide Area Network}
\newacro{mbb}[MBB]{Mobile Broadband}
\newacro{mfa}[MFA]{Multi-Factor Authentication}
\newacro{mcc}[MCC]{Mobile Cloud Computing}
\newacroplural{mcus}[MCUs]{Microcontroler Units}
\newacro{m2m}[M2M]{Machine-to-Machine}
\newacro{m2s}[M2S]{Machine-to-Service}
\newacro{mimo}[MIMO]{Multiple Input Multiple Output}
\newacro{mmimo}[mMIMO]{massive Multiple Input Multiple Output}
\newacro{ml}[ML]{Machine Learning}
\newacro{mulc}[mULC]{massive Ultra-Reliable Low-Latency Communication}
\newacro{mmtc}[MMTC]{massive Machine Type Communication}
\newacro{mmg}[MMG]{Mechanomyogram}
\newacro{multos}[MULTOS]{Multii-Application Smart Card Operating System}
\newacro{mux}[MUX]{Multiplexer}
\newacro{mnc}[MNC]{Mobile Network Code}
\newacro{me}[ME]{Mobile Environment}
\newacro{mpsoc}[MPSoC]{ Multiprocessor System on Chip}
\newacro{ngmn}[NGMN]{Next Generation Mobile Network}
\newacro{nic}[NIC]{Network Interface Controller}
\newacro{nist}[NIST]{National Institute of Standards and Technology}
\newacro{nn}[NN]{Neural Network}
\newacro{oath}[OATH]{Open Authentication}
\newacro{ocra}[OCRA]{\ac{oath} Challenge-Response Algorithm}
\newacro{ocsp}[OCSP]{Online Certificate Status Protocol}
\newacro{otp}[OTP]{One-Time Password}
\newacro{pap}[PAP]{Password-Authentication-Protocol}
\newacro{physec}[PhySec]{Physical Layer Security}
\newacro{pfs}[PFS]{Perfect Forward Secrecy}
\newacro{pin}[PIN]{Personal Identification Number}
\newacro{pkc}[PKC]{Public Key Cryptography}
\newacro{pki}[PKI]{Public Key Infrastructure}
\newacro{ppg}[PPG]{Photoplethysmography}
\newacro{prng}[PRNG]{Pseudo Random Number Generator}
\newacro{puf}[PUF]{Physically Unclonable Function}
\newacroplural{pufs}[PUFs]{Physically Unclonable Functions}
\newacro{pl}[PL]{Programmable Logic}
\newacro{ps}[PS]{Processing System}
\newacro{qos}[QoS]{Quality of Service}
\newacro{qr}[QR]{Quick Response}
\newacro{rat}[RAT]{Radio Access Technology}
\newacro{radius}[RADIUS]{Remote Authentication Dial-In User Service}
\newacro{ram}[RAM]{Random-Access Memory}
\newacro{ran}[RAN]{Radio Access Networks}
\newacro{rfid}[RFID]{Radio-Frequency Identification}
\newacro{ris}[RIS]{Reconfigurable Intelligent Surface}
\newacro{rng}[RNG]{Random Number Generator}
\newacro{ro}[RO]{Ring-Oscillator}
\newacro{rom}[ROM]{Read-Only Memory}
\newacro{rs}[RS]{Reed-Solomon}
\newacro{rsa}[RSA]{Rivest-Shamir-Adleman}
\newacro{rssi}[RSSI]{Received Signal Strength Indicator}
\newacro{rsrp}[RSRP]{Reference Signal Received Power}
\newacro{re}[RE]{Resource Elements}

\newacro{sdn}[SDN]{Software-Defined Network}
\newacro{sdr}[SDR]{Software-Defined Radio}
\newacro{seccos}[SECCOS]{Secure Chip Card Operating System}
\newacro{sip}[SIP]{Session Initiation Protocol}
\newacro{skg}[SKG]{Secret Key Generation}
\newacro{sram}[SRAM]{Static Random Access Memory}
\newacro{srs}[SRS]{Software Radio Systems}
\newacro{starcos}[STARCOS]{Smart Card Chip Operating System}
\newacro{sha}[SHA]{Secure Hash Algorithm}
\newacro{se}[SE]{Static Environment}
\newacro{snr}[SNR]{Signal-to-Noise-Ratio}
\newacro{tcg}[TCG]{Trusted Computing Group}
\newacro{tpm}[TPM]{Trusted Platform Module}
\newacro{tls}[TLS]{Transport Layer Security}
\newacro{trng}[TRNG]{True Random Number Generator}
\newacro{tsn}[TSN]{Time-Sensitve Networking}
\newacro{tofu}[TOFU]{Trust On First Use}
\newacro{tufu}[TUFU]{Trust Upon First Use}
\newacro{totp}[TOTP]{Time-based One-time Password Algorithm}
\newacro{uav}[UAV]{Unmanned Arial Vehicles}
\newacro{usb}[USB]{Universal Serial Bus}
\newacro{usrp}[USRP]{Universal Software Radio Peripheral}
\newacro{uhd}[UHD]{USRP Hardware Driver}
\newacro{usim}[USIM]{Universal Subscriber Identity Module}
\newacro{ue}[UE]{User Equipment}
\newacro{urllc}[URLLC]{Ultra-Reliable Low-Latency Communication}
\newacro{ulbc}[ULBC]{Ultra-Reliable Low-Latency Broadband Communication}
\newacro{umbb}[uMBB]{ubiquious Mobile Broadband}
\newacro{vlc}[VLC]{Visible Light Communicataion}
\newacro{warp}[WARP]{Wireless open-Access Research Platform}

%% file: chapter/introduction.tex
\label{sec:introduction}



With new emerging technologies in the field of communication and mobile systems, demands for smart and secure platforms are arising in real-world situations. Moreover, increasing the complexity of target applications with exclusive purposes leads to a wide spectrum of research focusing on constructing smart systems employing \acf{ai} solutions. For large-scale application models, the need for parallel computational capabilities with higher performance raises a lot of design challenges. For instance, \textit{Raina, Madhavan, and Ng} \cite{raina2009large} discuss the implementation of unsupervised learning methods over the modern graphics processor which scales 70 times faster than a dual-core \ac{cpu}. The development of Cloud computing technologies has enabled the provision of higher computational capabilities. Thus, the custom methodology for creating powerful \ac{ai} systems is using Cloud services due to their extensive computation power. However, the traditional Cloud solutions cannot provide the desired efficiency for daily increasing mobile systems \cite{ITU.2015}\cite{Jiang.2021}. As a result, a new computing architecture based on end devices (also known as IoT devices), edge devices, and Cloud has already been introduced and is used to improve the performance of systems that are extremely computationally dependent \cite{huang2017deep}.

Cloud servers could be used as a very powerful computing tool, particularly for running \ac{ai} inference systems; nevertheless, due to the significant data transmission between the end device and the Cloud, the issue of reliability arises. Moreover, the available bandwidth in a network is restricted, therefore expanding the size of the network may result in further bandwidth reductions concerning the increased data path. In addition, the data transfer within a network can compromise the data security and as a result, endanger the system security \cite{Lipps.2020a}.

A commonly used choice as a tool for edge computing is utilizing \ac{gpu} systems; this is due to the strong computation capabilities and also the high memory bandwidth of \ac{gpu} devices. However, these systems have a lot of issues for battery-dependent systems, especially for mobile applications. Two of the most important design parameters for every edge device are i) the energy efficiency and ii) the security of the design. \ac{gpu} systems are designed for graphical purposes and would not provide a fully reliable smart system. One considerable solution for implementing bandwidth-optimized, energy-efficient, and secure devices is using ARM-FPGA hybrid systems.  These systems consist of ARM processors as the heart of software processing which act as the \ac{ccu} alongside a dedicated hardware accelerator on \ac{fpga} blocks for boosting the desired \ac{ml} algorithm using the concept of parallelism. In 1992, \cite{sackinger1992application} demonstrated that a large neural network application with a single ANNA chip (\ac{anna}) enables it to operate with the speed advantage of 50 to 500 over conventional hardware evaluated with Floating-Point (FP) precision. 

In the current paper, an \ac{fpga} based edge solution is presented in comparison to the \ac{gpu} methods over the edge. The remainder of this work is organized as follows. Section \ref{sec:related_works} provides the current state of the art for the embedded \ac{fpga}'s using \acp{mpsoc}. The detailed structure of our proposed system with a dedicated hardware accelerator is presented in Section \ref{sec:our_work}. The implementation comparison of embedded \ac{fpga} and \ac{gpu} (Edge vs Cloud) are discussed in Section \ref{sec:comparison_edge_cloud}. An outline of future work and this work is concluded in Section \ref{sec:conlusion}.

%% file: chapter/related_works.tex
\label{sec:related_works}




First ideas of reducing network traffic by the use of edge-computation are provided by \cite{huang2017deep}. Therefore they tackled the problem of crowdsourced applications which are geo-distributed globally with highly heterogeneous crowdsourced data with the use of edge computing. 
Especially they want to combat the challenges in crowdsourced deep learning applications. 
They claim to be able to reduce the network traffic by 80\,\% and running time by 69\,\% in comparison to state-of-the-art cloud solutions. One of the big advantages is the improvement of network latency between end-user devices and edge servers compared to cloud servers. \cite{Li2018} and \cite{Li2020a} try to solve the problem of high computation-intensive \ac{dnn} based tasks on mobile devices with the development of a specific Framework, which claims to use a device-edge-synergy. Therefore they use algorithms to adaptively partition computations between device and edge. Furthermore, they are reducing computing latency via early exiting inference provided by the use of BranchyNet \cite{teerapittayanon2016branchynet} at the edge. 

\cite{xu2018efficient} examines latency and power optimizations of heavily computation-dependent applications on the client side with the use of \acp{fpga}. They investigate the efficient hardware implementation of Cellular Neural Networks (CeNNs) in \ac{fpga} with the help of different optimization algorithms for compressing image processing techniques. By using a so-called incremental quantization which includes parameter partition, parameter quantization, and re-training, they can quantize the numbers in CeNN templates to the power of two, so they are able to use the full potential of embedded \acp{fpga}. This solves the problem of the very high need for multiplications needed in CeNNs, which leads to a bottleneck in an FPGA because of the limited quantity of embedded multipliers one contains. By reducing the power of numbers to two the multiplications can be solved with the use of logic shifts, which can be achieved with logic elements and registers. 
In the represented work CNNs are used instead of CeNNs, moreover, it will be the Xilinx deep learning IP (Xilinx DPU) which delivers sufficient speed up for inference tasks.

Also \cite{Sharma2016} worked on more efficient ways of implementing efficient \acp{dnn} on \acp{fpga}. Therefore they developed a Framework named \textit{DNNWEAVER} which can automatically generate synthesizable accelerators for a given pair of DNN and \ac{fpga}. With performance tests, they compared three different FPGAs (Xilinx Zynq, Altera Stratix V, Altera Arria 10) against different many-core GPUs. They were able to achieve higher Performance-per-Watt against all tested many-core GPUs for the Xilinx Zynq and Altera Arria 10 FPGAs by using the framework synthesized DNNs. 

In \cite{he2021enabling} the authors went one step further and claim to create the first hybrid system of  \ac{gpu}-\acp{fpga} for training a network. Therefore they propose a new framework with a focus on effective \ac{dnn} training on GPU-FPGA-hybrids. They used different energy-efficient schemes for \ac{dnn} training procedures to execute individual training operations on either high-performance (GPUs) or power-efficient hardware (FPGAs). With that, they claim to be able to reduce the average power consumption by 44.3\,\%. 
All in all, it is to say the higher adaptability of \acp{fpga} and the higher more advanced processing power of embedded \acp{fpga} compared to normal GPUs result in lower cost and relatively lower power consumption which make them a serious threat to pure \ac{gpu} solutions still used in many classical cloud-based-solutions so far. 

%% file: chapter/our_work.tex
\label{sec:our_work}
This Section describes the proposed design with a dedicated hardware accelerator on \ac{fpga} using \ac{pl} system of Xilinx \ac{mpsoc} evaluation board (Xilinx ZCU102). The heart of this accelerator includes Xilinx \ac{dpu} \cite{xilinx:PG338} which is responsible for computing given inference tasks for a \ac{dnn} system. The main functionality of this system is to calculate massive series of Multiply–Accumulate operations (MAC) on network parameters of \acp{dnn}. On the other hand, the initialization task, as well as data coordination of a \ac{dnn}, will be done using ARM processors in the \ac{ps} of the \ac{mpsoc}.

\begin{figure}[!t]
    \centering
    \includegraphics[width=0.5\textwidth]{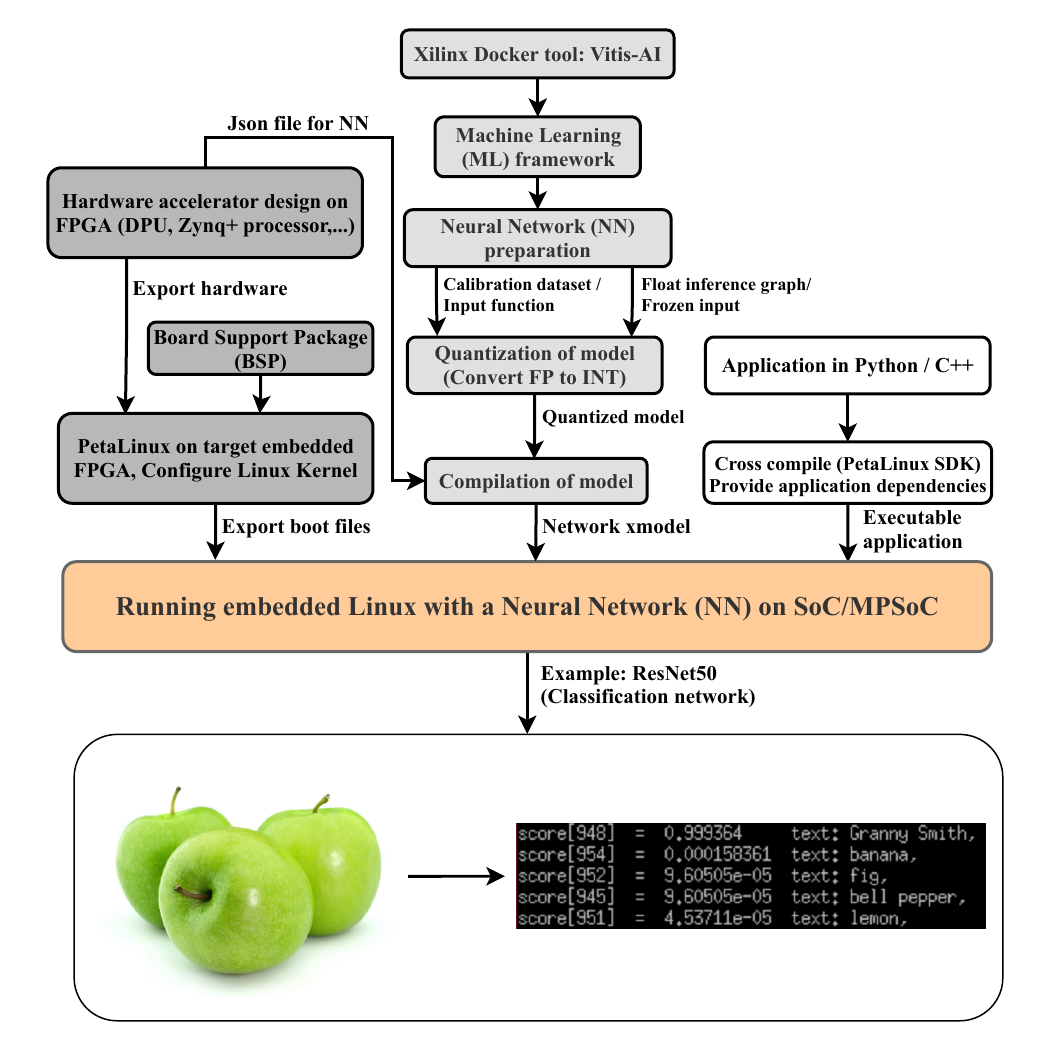}
    \caption{Overall workflow for implementing NNs on the target SoC/MPSoC board}
    \label{fig:workflow_xilinx}
\end{figure}

The overall workflow for preparing and implementing a \ac{nn} on a target embedded \ac{fpga} board (Xilinx SoC/MPSoC) is depicted in Figure~\ref{fig:workflow_xilinx}. This diagram consists of three sub-workflows. The most left path (depicted in dark gray) explains the required steps for preparing a dedicated hardware accelerator on \ac{fpga} logic gates using the Xilinx \ac{dpu} soft IP. This path also includes all the desired pre-processing at the hardware level. Then using the implemented design in the \ac{pl} side, computational processes of a \ac{nn}, which are mainly multiplication and addition functions, are performed on the \ac{fpga} side. The design strategy behind the DPU IP is utilizing \ac{dsp} tiles for conducting MAC (Multiply-Accumulate) operations. 

The middle sub-workflow (depicted in light grey) describes the required steps for converting the desired NN from its initial form to an equivalent \ac{fpga} compatible form. This step involves model quantization (converting floating-based weights and biases to equivalent integer numbers). The quantization step is very important not only because it reduces the required storage memory but also optimizes the required bandwidth for the data transmission between the off-chip memory (DDR memory connected to the \ac{ps} side) and the hardware accelerator on the \ac{fpga} side. After the quantization step, the graph-based NN must be converted to a binary form which is executable on logic gates of the PL subsystem. In this step, the pre-quantized model is fed into the Xilinx compiler for generating the xmodel file of the initial NN. 

For controlling the data transmission among various parts of the SoC/MPSoC as well as for servicing the interrupts, a user-defined Linux application should be used. This part is shown in the most right sub-workflow (depicted in white). Based on the requirements of the project, there may be a need for cross-compiling the given application (C++ application). 

Finally, by booting the embedded Linux (PetaLinux) with a given NN on the target board, further experiments are possible. In this case, the classic ResNet50 \cite{he2016deep} DNN is used which is a pre-trained classification network on $224\times224\times3$ ($224\times224$ pixels with 3 color channels) images from ImageNet dataset \cite{deng2009imagenet}. The current version of the ResNet50 network can classify 1000 different classes based on given input images.

\begin{figure*}[h]
  \centering
  \includegraphics[scale=0.36]{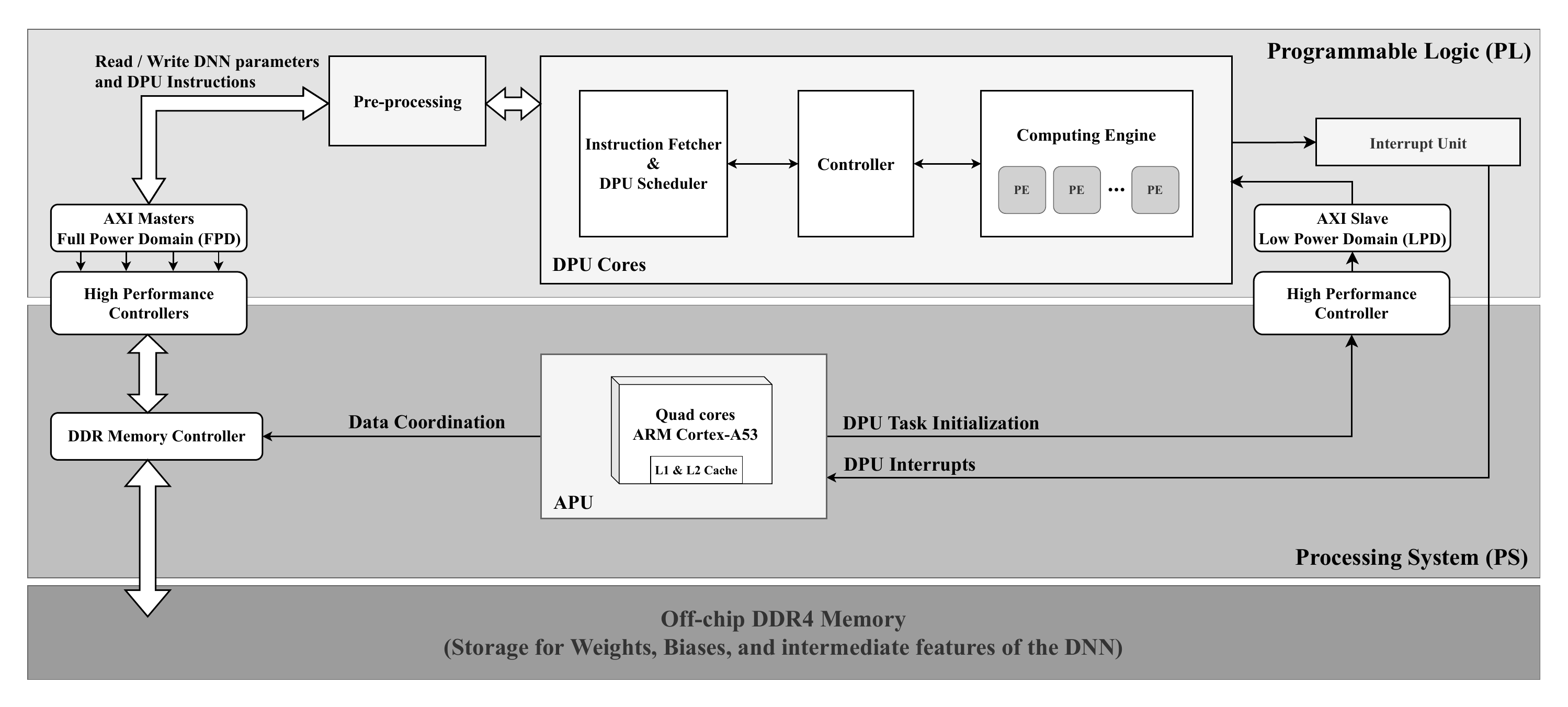}
  \caption{The hardware accelerator design for implementing a DNN on the target MPSoC}
  \label{fig:hw_acc}
\end{figure*}

The overall design scheme for the hardware accelerator used in this paper is depicted in Figure~\ref{fig:hw_acc}. This design includes DPU core(s) used for the inference computation of the target NN. The DPU core is responsible for the computation of pixels in the feature map of input images. The number of Processing Elements (PEs) inside the convolution engine is equivalent to the pixel parallelism of the feature map. The implemented NN parameters are stored in off-chip memory (DDR4 memory) and will be transmitted to the PL system with the Full Power Domain (FPD) AXI connections. For boosting the performance of the accelerator, connections between hardware logic in FPGA and off-chip memory are built through a Direct Memory Access (DMA) system. In this case, PL can transmit data directly to or from external memory. The ARM processors in the APU subsystem are responsible for data coordination as well as required initialization for given tasks to DPU core(s).

%% file: chapter/comparison_edge_cloud.tex
\label{sec:comparison_edge_cloud}






The \ac{dnn} implementations over the Edge and Cloud are presented in this Section. Moreover, the detailed description of the environment setup for used Edge and Cloud components, \ac{fpga} and \ac{gpu} features, and the model used for the execution of the \ac{dnn} are mentioned in related parts.

\subsection{Deep Neural Network Implementation over Edge}

The implementation details of a \ac{dnn} on an edge-based device are described in this Subsection. \\

\subsubsection{Physical Infrastructure}
 The edge device used in this set-up is the ZCU102 evaluation board which is an embedded \ac{fpga} from Xilinx Inc. This board is considered as a \ac{mpsoc} with the Zynq UltraScale+ architecture. However, the overall workflow presented in this paper can be replicated in any Zynq and Zynq UltraScale+ based evaluation or custom boards (for example, Zynq-7000 ARM-FPGA SoC boards). The main reason for selecting a Zynq UltraScale+ system is the scalable capability of MPSoC family boards, which is an important factor for creating smart networks. Moreover, ZCU102 provides more logic resources in \ac{fpga} (including DSP blocks, which are the main computational units) for performing required computations of the inference graph of DNNs. Another reason for selecting the Zynq UltraScale+ over the Zynq architecture is the possibility of implementing the SoftMax activation function at the hardware level. The hardware implementation of this activation function can boost the performance of the system much more than the equivalent function at the software level. The ZCU102 consists of ARM Cortex processors in the Application Processing Unit (APU) and  Real-time Processing Unit (RPU) sub-systems which are implemented in the \ac{ps}. Moreover, a mobile \ac{gpu} is encompassed in the PS side of the board. The system specifications of Xilinx ZCU102 MPSoC are described in Table \ref{tab:embedded_FPGA_Specifications}.

\begin{table}[h!]
\centering
\caption{Embedded FPGA Hardware Specification}
\begin{tabular}{||l l||}
 \hline
 Processing System (PS): & \\ [1ex] 
 \hline \hline
 APU & Quad-core ARM Cortex-A53\\ 
 \hline
 RPU & Dual-core ARM Cortex-R5\\
 \hline
  GPU & Mali-400\\
 \hline\hline
  Programmable Logic (PL): & \\ [1ex] 
 \hline \hline
 System Logic Cells & 600 K\\ 
 \hline
 Block Ram (BRAM) & 32.1 MB\\
 \hline
 DSP Slices & 2520\\
 \hline\hline
 Memory types: & \\ [1ex] 
 \hline \hline
 PS-Side DDR4 & 4GB 64-bit\\ 
 \hline
 PL-Side DDR4 & 512MB 16-bit\\ 
 \hline
 \end{tabular}
 \label{tab:embedded_FPGA_Specifications}
\end{table}

\subsubsection{DNN Implementation over Edge device} 
The implementation results for ResNet50 DNN on the Xilinx ZCU102 MPSoC board are shown in Table \ref{tab:mpsoc_latency}. In this work, several test scenarios are implemented on the target board using different sizes and numbers of DPU cores. The achieved throughput and power consumption of the system depend on the actual requirements of a project; however, in these test scenarios assumed a power budget of around 10 to 20 Watts is available and also the target throughput of the system is around 100 to 200 Images per Second (Img/S). Achieving values higher than the proposed range is possible by utilizing more logic resources on \ac{fpga}, or using more expensive and advanced embedded FPGAs, or providing a bigger power supply with a proper cooling system. One important design parameter for these experiments is the Convolution architectures of the DPU core. According to the Xilinx documentation \cite{xilinx:PG338}, this design parameter determines the number of operations the DPU core can perform in one Clock Cycle (CC). For example, 1/B4096 architecture equals 1 DPU core capable of performing 4096 operations in one CC. The bigger the Convolution architectures means higher the logic resources used for creating that DPU core. The maximum number of implementable DPUs is limited due to the high \ac{dsp} block usage in the internal structure of this AI unit. The current version of this soft IP core supports up to 4 homogeneous cores. Considering the results from these experiments, the direct relationship between the power consumption and the DPU core size is considerable. The number of test images used for each scenario is equal to 2000 images.

\begin{table}[h!]
\centering
\caption{Power, Latency and Throughput of the Edge device}
\begin{tabular}{||c c c c||} 
 \hline
 DPU  Cores/Arch & Power(W) & Latency(S) & Throughput(Img/S)\\ [1ex] 
 \hline\hline
 1/B4096 & 9.97 & 22.59 & 88.5\\ 
 \hline
 2/B4096 & 16.29 & 12.40 & 161.2\\
 \hline
 3/B4096 & 22.86 & 9.73 & 205.4\\
 \hline
 4/B2304 & 21.98 & 11.68 & 171.1\\
 \hline
 \end{tabular}
 \label{tab:mpsoc_latency}
\end{table}

\subsection{Deep Neural Network Implementation over Cloud}
A detailed description of \ac{dnn} implementation constructed on a Cloud-based \ac{gpu} cluster infrastructure is discussed. \\

\subsubsection{Physical Infrastructure}
The experimental setup consists of a Cloud Computing cluster with \ac{gpu} servers namely GTX1080Ti and RTX A6000. Table \ref{tab:GPU_Specifications} provides the overall \ac{gpu} specifications used in the implementation. Slurm \cite{Slurm:Online} is used for managing compute resources, scheduling jobs, and execution on worker nodes, it provides fault-tolerant and highly scalable cluster management.

\begin{table}[h!]
\centering
\caption{GPU Cluster Hardware Specification}
\begin{tabular}{||l l l||}
 \hline
 Specifications: & GTX1080Ti& RTX A6000 \\ [1ex] 
 \hline \hline
 Architecture  & Pascal & Ampere\\ 
 \hline
 GPU Memory(GB)  & 11 & 48\\
 \hline
 GPU per node & 8 & 8\\
 \hline
 CPU per GPU  & 5-9 & 12\\ 
 \hline
 \end{tabular}
 \label{tab:GPU_Specifications}
\end{table}

\subsubsection{DNN Implementation over GPU Cluster}

With the deployed cluster environment, \ac{dnn} is constructed using a containerized environment using enroot, administrated, and monitored through the Slurm cluster. The framework is built using Keras with TensorFlow \cite{tensorflow2015} backend to run the image identifier model using ResNet50. The ResNet50 model employed in this study is the same pre-trained model used in the Edge implementation. \ac{gpu}s were subjected to iterative image prediction analysis and throughput as well as the total power consumption are recorded in Table \ref{tab:gpu_throughput}. The implementation is tested over 2 different \ac{gpu} servers working with  \ac{gpu} in sets of 2 and 4 along with 2 \ac{cpu}s. The CPU used alongside the GTX1080Ti \ac{gpu} is the Intel(R) Xeon(R) \ac{cpu} E5-2630 v4 and the \ac{cpu} used with the RTXA6000 \ac{gpu} is the AMD(R) EPYC 7F72 24-Core Processor. The overall latency is the total time taken during the resource allocation, containerized environment creation, initialization, and final image prediction. \ac{gpu}s are allocated with a batch size of 8, to obtain a minimalistic comparison to the edge device. With the increased batch size, high throughput can be achieved with higher \ac{gpu} utilization by compromising to operate with huge power. Thus, the experiment performed uses lower \ac{gpu} capacity, providing a lower throughput. The total number of test images for each experiment is equal to 2000.

Firstly, the system is tested with resources of 2 \ac{gpu}s and 2 \ac{cpu}s are allocated and tested on servers RTXA6000 and GTX1080Ti, and throughput obtained are 180 and 175 with the latency of 65s and 46s. Then, the allocated number of \ac{gpu}s was increased to 4, and throughputs of 187 and 137 were obtained. One important fact about the power consumption of these systems is the \ac{gpu} utilization, the test scenario with 2/RTXA6000 has higher utilization than 4/RTXA6000 and this is the reason for the higher power consumption.

\begin{table}[h!]
\centering
\caption{Power, Latency and Throughput of the GPU Cluster}
\begin{tabular}{||c c c c||} 
 \hline
 Num/GPU & Power(W) & Latency(S) & Throughput(Img/S)\\ [0.5ex] 
 \hline\hline
 2/RTXA6000 & 215.8 & 65 & 180\\ 
 \hline
 4/RTXA6000 & 174.4 & 83 & 187\\
 \hline
 2/GTX1080Ti & 95.9 & 46 & 175\\
 \hline
 4/GTX1080Ti & 138.4 & 68 & 137\\
 \hline
 \end{tabular}
 \label{tab:gpu_throughput}
\end{table}

\subsection{Comparison results of DNN Implementation over Edge versus GPU Cluster}

\begin{figure}[h!]
    \centering
     \includegraphics[width=0.5\textwidth]{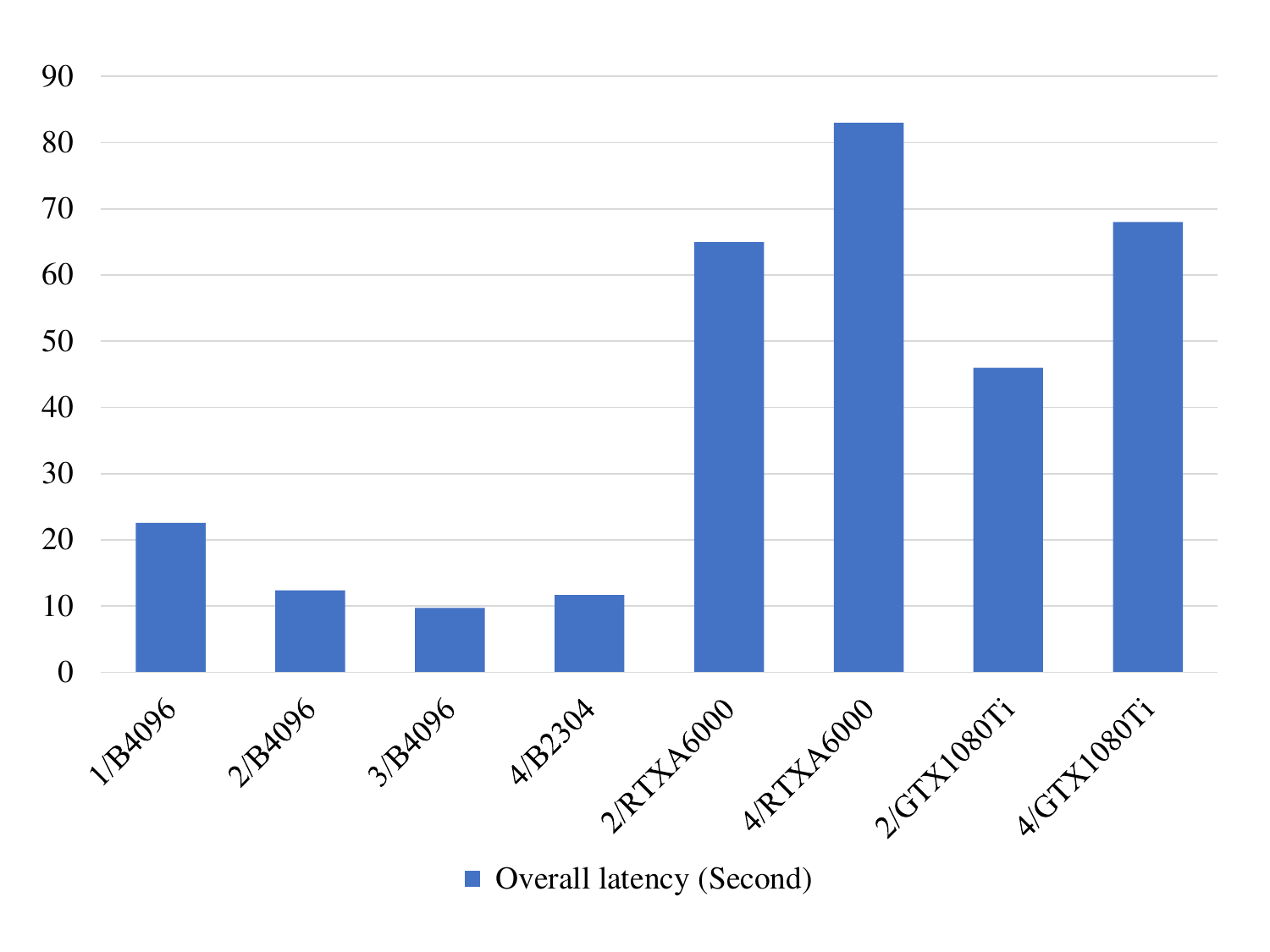}
     \caption{Latency comparison between the Edge and Cloud implementations}
    \label{img:edge_cloud_latency} 
\end{figure}

\begin{figure}[h!]
    \centering
     \includegraphics[width=0.5\textwidth]{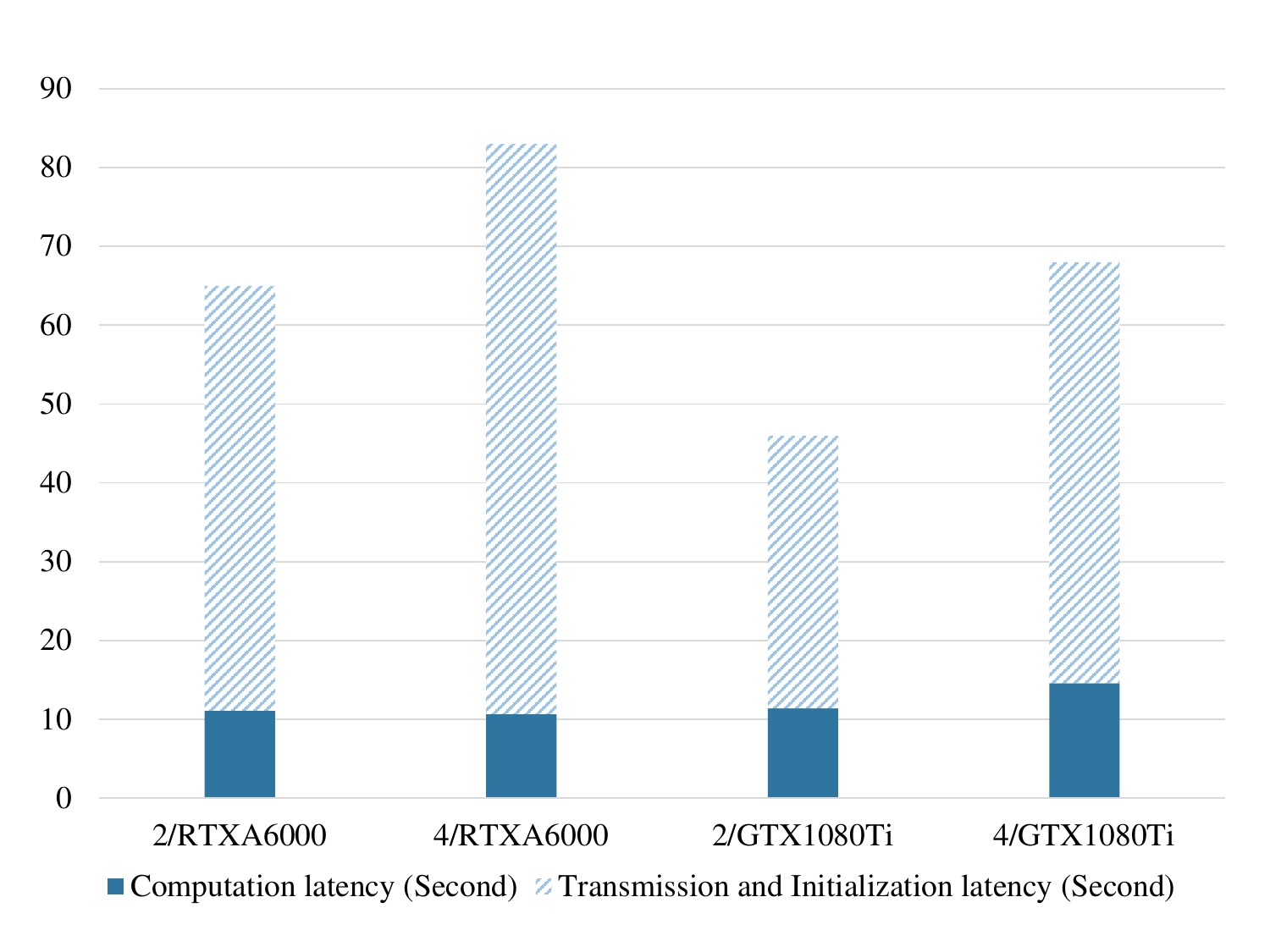}
     \caption{Latency sources in Cloud implementations}
    \label{img:detailed_cloud_latency} 
\end{figure}

Considering the obtained results from implementing our target DNN (Classic ResNet50) over Edge (embedded \ac{fpga} device) versus \ac{gpu} Cluster (as the Cloud tool) shows drastically lower latency using Edge solution. Figure \ref{img:edge_cloud_latency} shows the latency comparison of these two methodologies. The main reason for lower latency using embedded \ac{fpga}s is the negligible additional overhead for data transmission and initialization in the target board. Therefore, in an Edge device, the latency value equals the time it takes to compute the input data on the board (computation latency). However, in a Cloud implementation, there are other latency sources for data transmission over the Cloud, creation of new instances, and initialization of values. In this work, all the additional latency sources in the Cloud are called "Transmission and initialization latency". Our experiments show high flexibility and low latency of Edge devices using hardware accelerators built on \ac{fpga} logical units. Figure \ref{img:detailed_cloud_latency} shows the detailed latency sources using the Cloud tool.

Moreover, using embedded \ac{fpga}s is more energy-efficient than \ac{gpu}s and this design parameter is a decisive point for mobile projects with limited power sources. The noticeable point is, \ac{gpu} clusters are initially able to achieve much higher throughput than our test scenarios, but this is equal to a much higher power consumption of the system. Increasing the batch size leads to higher \ac{gpu} utilization, and this will result in higher throughput. For instance, in our experiment with increasing the batch size from 8 to 200, the throughput will jump from 137 Img/S to 751.49 Img/S in the 4/GTX1080Ti; however, the total power consumption changes from 138.4 Watts to 747 Watts. With replicating the same experiment using the 4/RTXA6000 scheme, the throughput jumps from 187 Img/S to 1886.13 Img/S, and at the same time overall consumed power changes from 174.4 Watts to 899 Watts. Considering the power budget as well as the practically required throughput of the DNN, our experiments show better energy efficiency in all the test cases. 

%% file: chapter/conclusion.tex
\label{sec:conlusion}
In this paper, we proposed a \acl{dnn} implementation strategy on Edge devices using a\acl{mpsoc}. Our experiments show that using Edge devices for \acl{ai} inferences has superiority in comparison to the Cloud implementation of the same network not only in the latency optimization but also in the energy efficiency of the system. One of the main components used in the hardware accelerator of this work is the Xilinx dedicated soft IP core for deep learning purposes (Xilinx DPU). The presented system can boost the computation process of DNNs using \acl{dsp} tiles and logic gates on the FPGA side as well as using ARM processing for controlling and coordination purposes. In future research, the possibility of implementing multiple \ac{dnn}s on \ac{mpsoc}s will be examined to observe the effect of co-implementation of \ac{dnn}s. Moreover, another possibility for future work is using the Xilinx Versal boards \cite{vissers2019versal} as the next generation of AI cores with dedicated AI engines instead of soft IP cores. 